\documentclass[conference]{IEEEtran}
\IEEEoverridecommandlockouts
\usepackage{cite}
\usepackage{amsmath,amssymb,amsfonts}
\usepackage{algorithmic}
\usepackage{graphicx}
\usepackage{textcomp}
\usepackage{xcolor}

\usepackage[nolist,nohyperlinks]{acronym}
\usepackage[english]{babel}
\usepackage{hyperref} 
\usepackage{cleveref}
\def\BibTeX{{\rm B\kern-.05em{\sc i\kern-.025em b}\kern-.08em
    T\kern-.1667em\lower.7ex\hbox{E}\kern-.125emX}}

   \begin{acronym}[MMMMMMMMM]
      \acro{I4.0}{industry 4.0}
      \acro{IoT}{internet of things}
      \acro{CPS}{cyber physical systems}
      \acro{IIoT}{industrial internet of things}
      \acro{PLC}{programmable logic controllers}
      \acro{OT}{operational technology}
      \acro{IT}{information technology}
      \acro{SCADA}{supervisory control and data acquisition}
      \acro{ICS}{industrial control systems}
      \acro{IACS}{industrial automation and control system}
      \acro{DSR}{design science research}
      \acro{IEC}{International Electrotechnical Commission}
      \acro{ISA}{International Society of Automation}
      \acro{OSM}{organizational security measure}
      \acro{SLM}{security level measure}
      \acro{AO}{asset owner}
      \acro{SI}{integration service provider}
      \acro{SM}{maintenance service provider}
      \acro{SL}{security level}
      \acro{SPR}{security program rating}
      \acro{SR}{security requirement}
      \acro{SR-P}{security requirement portfolio}
      \acro{FR}{foundational requirement}
      \acro{SL-T}{security level target}
      \acro{SL-A}{security level achieved}
      \acro{SPR-A}{security program rating achieved}
      \acro{IAC}{identification and authentication control}
      \acro{DoS}{denial of service}
      \acro{EGM}{evidence gathering mechanism}
      \acro{MIP}{measurable indicator point}
      \acro{MSI}{measurable security indicator}
      \acro{IDS}{intrusion detection system}
      \acro{HIDS}{host-based IDS}
      \acro{NIDS}{network-based IDS}
      \acro{BPM}{business process management}
      \acro{BPMN}{business process management and notation}
      \acro{UML}{unified modeling language}
      \acro{CMS}{compliance monitoring system}
      \acro{RE}{requirement enhancement}
      \acro{FTP}{file transfer protocol}
      \acro{SFTP}{secure file transfer protocol}
      \acro{H2H}{human-2-human}
      \acro{P2P}{person-2-person}
      \acro{H2M}{human-2-machine}
      \acro{M2M}{machine-2-machine}
      \acro{HMI}{human-machine-interface}
      \acro{MAC}{media access control}
      \acro{NetBIOS}{network basic input /output system}
      \acro{EPC}{Electronic Product Code}
      \acro{Ucode}{ubiquitous code}
      \acro{MFA}{Multi-Factor authentication}
      \acro{TLS}{Transport Layer Security}
      \acro{SSL}{Soft Secure Layer}
      \acro{DTLS}{Datagram TLS}
      \acro{PKI}{Public-Key-Infrastructure}
      \acro{HTTP}{Hypertext Transfer Protocol}
      \acro{CoAP}{Constrained Application Protocol}
      \acro{MQTT}{Message Queue Telemetry Transport}
      \acro{XMPP}{Extensible Messaging or Presence Protocol}
      \acro{OPCUA}{Open Platform Communication Unified Architecture}
      \acro{EAP}{Extensible Authentication Protocol}
      \acro{LDAP}{Lightweight Directory Access Protocol}
      \acro{CA}{certification authority}
      \acro{FAT/SAT}{factory- or site-acceptance tests}
      \acro{RPC}{Remote Procedure Call}
      \acro{BPD}{boundary protection devices}
      \acro{SIEM}{security incident and event management}
      \acro{TSV}{tab-separated values}
   \end{acronym}

\begin{document}

\title{Market Research on IIoT Standard Compliance Monitoring Providers and deriving Attributes for IIoT Compliance Monitoring\\
}

\author{\IEEEauthorblockN{1\textsuperscript{st} Oberhofer Daniel}
\IEEEauthorblockA{
\textit{University of Regensburg}\\
Regensburg, Germany \\
daniel.oberhofer@informatik.uni-regensburg.de}
\and
\IEEEauthorblockN{2\textsuperscript{nd} Hornsteiner Markus}
\IEEEauthorblockA{\textit{University of Regensburg}\\
Regensburg, Germany \\
markus.hornsteiner@informatik.uni-regensburg.de}
\and
\IEEEauthorblockN{3\textsuperscript{rd} Schönig Stefan}
\IEEEauthorblockA{\textit{University of Regensburg}\\
Regensburg, Germany \\
stefan.schoenig@informatik.uni-regensburg.de}
}

\maketitle

\begin{abstract}
Adapting security architectures to common standards like IEC 62443 or ISO 27000 in the Industrial Internet of Things (IIoT) involves complex processes and compliance reports. Automatic monitoring of compliance status would enhance this process. Despite limited research, practical applications exist. This paper conducts a market study on providers implementing IEC 62443 in IIoT, aiming to formulate a catalog of monitorable attributes aligned with the standard. The study reveals challenges, such as a lack of formal separation in security architectures, limiting visibility. Despite these challenges, practical implementations share commonalities, providing insights into viable monitoring properties. The research serves as a crucial entry point into developing a comprehensive catalog of monitorable attributes for IEC 62443 standards in IIoT.

Aligned with the IEC 62443 SR catalog of document 3-3 \cite{iec_62443}, monitorable attributes are derived based on current research about IIoT security and Expert Knowledge. The provided tables serve as an exemplary extract, not exhaustive, defining three types of attributes based on their origin of creation.
\end{abstract}


\section{Introduction}
Adapting a security architecture to the \acp{SR} of common security standards like the IEC 62443 or the ISO 27000 series is a complex process involving multiple actors, processes and compliance reports. An automatic monitoring of the actual compliance status of an \ac{IIoT} system would be beneficial to the overall compliance process. To establish a comprehensive solution for such a compliance monitoring system, defining monitorable attributes is imperative to enable a technological implementation. Despite limited existing research in this domain, practical applications are already in use. This paper serves as a foundational exploration into this realm by conducting a market study in \cref{sec:market_research}, specifically examining various providers implementing the IEC 62443 standard in the context of \ac{IIoT}. Based on the market research, we formulate a catalog of monitorable attributes aligned with the IEC 62443 standard, discussed in \cref{sec:FR1}. Overall can the results of this market study be adapted to other standards, if their requirement structure provides similarity.

\section{Market Research on IIoT Standard Compliance Monitoring Providers}
\label{sec:market_research}
The practical market study involves an examination of companies that integrate compliance monitoring systems as part of their broader security architecture for IIoT. It is observed that many such solutions embed compliance monitoring and verification within the overarching security architecture, lacking a formal separation of the according processes. This integration approach blurs the visibility of the boundary between compliance related monitoring efforts and the overall security product. This makes differentiating in between companies focusing on implementing measures to fulfill the \acp{SR} and companies providing products checking the compliance, difficult.
Additionally, the pool of publicly available information on these products is limited, as there are no open-source tools implementing compliance related monitoring systems. However, existing approaches used in practice can offer valuable insights for constructing a portfolio of monitorable attributes. Despite the challenges, these practical implementations share commonalities, providing a basis to enhance the understanding of viable monitoring properties in real-world scenarios.

Consequently, this paper initiates a market study focusing on companies implementing IEC 62443 compliance monitoring, either partially or holistically. The primary aim is to identify similarities in monitoring approaches across different providers, thereby improving the understanding of practical attributes applicable in compliance monitoring. This research serves as a crucial entry point into the development of a comprehensive catalog of monitorable attributes aligned with IEC 62443 standards in IIoT environments.

\subsection{Methods}
The market research focused on publicly available information, for example provided by free white papers or tool documentations and also included providers focusing on the implementation of \acp{SLM} in the context of the IEC 62443. Table \ref{tab:mr_data} gives an overview over the investigated material \footnote{Sources provided in the appendix}. It provides the names of companies and the focus of their products or business model. The column describing the focus of the companies value proposition is necessary to differentiate in between products, increasing the transparency of which sources were finally included in the evaluation of the material. Although this categorization is somewhat subjective and does not show distinct borders, as described above. The categories of investigated products differ in between companies providing simple \ac{SLM} solutions, whole architectures, and companies partially including compliance monitoring in their products. Products implementing compliance aspects must include the functionality to specifically measure compliance and also mention compliance as being an intentional part. But even the investigated products within this class do not classify as standard compliance monitoring systems in our conception of the field. This is the result of not just checking the compliance but also implementing \ac{SLM}s like e.g., providing the functionality of a central certification authority. In practice though it might be an advantage to unify compliance measurement with \ac{SLM} implementations, as this enables new sources of monitorable attributes. In the last column of table \ref{tab:mr_data} the category of the investigated material is documented to increase the transparency of the market study. Most of the information gathered was provided by publicly available white papers with the intention of introducing potential customers to the advertised products or services. As this publication is an independent study, all of the companies material was treated equally and none of the companies were contacted. In addition to the mentioned categories of products was the documentation of the cdp studio software also included in the study. Cdp studio is a development environment used to implement the software of \ac{IACS} systems. It was included in the market study, as they describe an example for a \ac{SLM} for each of the requirements in the open access documentation of the program helping with the following attribute derivation. Each of the investigated sources is provided as a reference noted that the source material could be replaced, relocated, or deleted due to newer product versions or deprecated technology.
\begin{table}[h]
    \centering
    \begin{tabular}{|c|c|c|}
        \hline
        \textbf{Company} & \textbf{Focus} &  \textbf{Source} \\
        \hline
        Armis & Compliance \& SLM  & White paper\\
        Endian & Architecture  & White paper\\
        Schneider & Architecture  & White paper \\
        Cisco & Architecture & White paper \\
        ScadaFence & Compliance  & White paper  \\
        Applied Risk & Consulting  & Case Study \\
        NXP & Consulting & Article  \\
        Industrial Defender & SLM \& Compliance & E-Book  \\
        Fortra & Architecture &  Data sheet \\
        Radiflow & Compliance \& SLM &  White paper \\
        Baracuda & SLM &  White paper  \\
        cdp & SLM & Documentation \\
        Fortinet & Architecture & White paper \\
        ForeScout & Compliance  & White paper \\
        Rhebo & Compliance  & White paper\\
        \hline
    \end{tabular}
    \caption{An overview over the in the market study analyzed products in the field of Security Standard Compliance Monitoring}
    \label{tab:mr_data}
\end{table}
\\ In summary were the products of 15 companies engaging in the topic of IEC 62443 analyzed. As a result only three of the companies analyzed focus on the compliance as their main goal, three more include compliance monitoring in their implementation of SLMs, five provide a whole standard compliance architecture including multiple separate products representing SLMs
, and two describe only the implementation of a SLM as their product. In addition two companies focus on the consulting side of the IEC 62443 standard compliance providing information about the certification process or presenting historical use cases. The results of the investigation concerning consulting services or products implementing whole architectures was not considered within the results presented in the following section. In order to get included in the evaluation of the market study a source had to explicitly mention compliance as a focus of implementation or as part of their business model.

\subsection{Results}
In the following are the results of the market study presented. The main focus of the investigation was to find commonalities in monitorable attributes or characteristics. Such a monitorable attribute or characteristic is defined as follows. An attribute is monitorable and therefore included in the results, if it is \textit{possible to measure its current status by the means of passive or active monitoring approaches}. This includes network-based attributes e.g., the protocol type or attributes requiring component specific knowledge e.g., local firewall configuration files. By comparing the products used in practice can a starting point for the more detailed attribute derivation be created. \\ Table \ref{tab:mr_results} shows the results of the market study in form of the detected monitoring characteristics in combination with the actual attributes used by the products to measures their status. As described before do some products implement more functionalities than provided by just a compliance monitoring system. This leads to the conclusion that not every attribute might be able to get used in an actual monitoring context. 
\begin{table}[h]
\centering
    \begin{tabular}{|c|l|}
        \hline
        \textbf{Monitoring Characteristic} & \textbf{Attributes} \\
        \hline
        Unknown protocols & Protocol type, protocol version, context \\
        Unknown communication & human user, protocol type, wireless external  \\
        Unknown Software Process  & Process-ID, device  \\
        Abnormal Behaviour & Amount or timing of communications \\
        Weak Encryption & encryption suites, keys  \\
        Insecure protocol & isHuman, wireless, external \\
        Insecure Credentials & Default or weak passwords  \\
        Session Management & Session-Timing or -ID   \\
        Traffic accumulation & Amount of authentication-attempts  \\
        Data Integrity & Error codes, session-timing, packet fragments  \\
        Network Management Incidents & Protocol type (Mangement protocol)  \\
        \hline
    \end{tabular}
    \caption{The results of the market study presenting valuable attributes that can be used to design a Compliance Monitoring System}
    \label{tab:mr_results}
\end{table}
\\ The first similarity in between the products is the monitoring of abnormal behavior of the \ac{IACS} process or the more detailed versions of that, which includes the monitoring for either unknown protocols, unknown communications taking place or unknown software processes. In the same scope is the monitoring of abnormal behavior mentioned in multiple documents, which focuses on the monitoring of the communication timing or the amount of transferred packets in regard to a known process specification. As most of the investigated products focus on the monitoring of the factor unknown and the general abnormal behavior of the processes, it can be noted that a theoretical definition of such a \ac{CMS} has to include the analysis of abnormal behavior as a foundational principle. An approach using specification-based IDS technologies allows the adaption of abnormal behavior analysis \cite{Zhou_2018}. Focusing on the monitorable attributes in that context do the results show that the protocol type holds great information value. This can be elevated for the monitoring of multiple characteristics. For example can the type of protocol be used to identify unknown communications in between two rightful users. As both the sender and the receiver identifiers might be right for the actual sequence of the process, though matched with the wrong protocol type, it is possible for the monitoring system to make the according assumptions. Another group of characteristics identified among the products is monitoring the strength of discovered security measures. This can be of potential for the \ac{CMS}, as there are requirements in the IEC 62443 like e.g., concerning the strength of encryption. As the first security strength-related characteristic do the sources mention insecure protocols with the monitored attributes of protocol type, version, and the context of the protocol being used. An example for this can be the transfer of a file into the control system via the \ac{FTP} in a context where the \ac{SFTP} should have been used. In such a case there would be the wrong protocol type used during the wrong process sequence. Another strength-related characteristic is the monitoring of traffic for weak encryption. Here do the analyzed products claim to identify the encryption suites and the strength of the keys used in the monitored network traffic. Even though the strength of the security measures was identified as a group of monitoring characteristics, is the description of actual technological implementations of these characteristics in form of attributes not given by the documents. The identification of the encryption suites might not be possible in the context of a \ac{CMS}. This leads to the assumption that the investigated products elevate the capabilities of their holistic security architecture  and the according information base to check compliance. At last do the products describe the measurement of the strength of credentials used for authentication as the last security strength-based characteristic. Similar to the encryption ciphers does this monitoring characteristic require more sufficient monitoring capabilities as provided by being part of the whole security solution. Another characteristic identified is the monitoring of data integrity. The analyzed products monitor this requirement by identifying integrity mismatches within the network traffic for example by detecting network-based error codes, fragmented packets or odd network flow timing. \\ The collection of similar characteristics in between the analyzed products can be seen as a starting point for a more detailed collection of monitorable attributes. Deriving such a collection requires great domain knowledge provided either by looking at practice as described or by looking at state-of-the-art surveys in the domain of \ac{IIoT} security.

\section{Deriving Attributes for IIoT Compliance Monitoring}
\label{sec:FR1}
Aligned with the IEC 62443 \ac{SR} catalog of document 3-3, monitorable attributes are derived based on current research about IIoT security and Expert Knowledge. The provided tables are not exhaustive, serving as an exemplary extract of monitorable attributes that could serve as a compendium to implement an actual monitoring solution. Some attributes may be feasible only for monitoring in specific scenarios, while others may emerge from specific system properties and are not listed here. In general, this section defines three types of attributes based on their origin of creation, which are presented in the following.
\begin{itemize}
    \item Traffic attributes: They are based on monitorable properties within the gathered traffic evidence or properties collected by active components. Examples for these kinds of attributes include the protocol type or configuration files.
    \item Logical attributes: They can be deducted logically by combining the gathered evidence with additional available information. An example for this would be an attribute classifying an entity as a human participant by comparing the monitored identifier with an available list of human participants to set a boolean value for the logical attribute.
    \item Manual attributes: These kind of attributes cannot be monitored or inferred logically based on the network traffic or by active components and must be set manually by using other means of compliance checking. Examples for this include the investigation of hardware-based requirements by an expert
\end{itemize}
Within the provided tables in the beginning of each subsection, there are attributes hinting to include a boolean value. Such attributes include either the value true if they are fulfilled or false if they are not fulfilled. Depending on the category of attributes could these boolean values be inferred logically or set manually in some cases. Especially the manual attributes are depicted as boolean based attributes. Even though the adaption of such attributes to the monitoring specifications shows a restricted usage, they are listed in the tables. In general are these attributes necessary to depict a verifiable value for the monitoring solution and should be implemented in some form and place. 

\subsection{FR 1: Identification and Authentication Control}
\label{FR:1}
The \acf{IAC} of entities in the IIoT is an important part of \ac{IIoT} security and can be seen as difficult to implement  \cite{lipps2021dance}. \ac{IIoT} devices might have limited computational resources or energy capabilities, which makes using complex protocols, encryption, and authentication measures not feasible in some scenarios \cite{Yu_2019}. As part of this process in \ac{IIoT} systems, the communication categories of \ac{H2H} and \ac{H2M} evolved with the increasing connectivity to now include a growing amount of \ac{M2M} communications \cite{Tan_2021}. The IEC 62443 combines identification and authentication into one \ac{FR} even though it is not considered the same. Identification describes the differentiation between entities in the system e.g., with an unique identifier, whereas authentication describes the verification of the given identity to the system \cite{qureshi2018survey}. Authentication of entities in \ac{IIoT} systems is based on the factors knowledge, inherence or possession \cite{lipps2021dance}. Possession includes physical or digital objects, knowledge includes the possession of secret, and inherence includes attributes of the entity to authenticate \cite{lipps2021dance}. To monitor the fulfillment of the in FR 1 included requirements, attributes have to be identified and assigned, based on characteristics of used authentication measures and identification solutions. Table \ref{tab:FR1} shows the derived attributes for the first \ac{FR} attached to their relevant \ac{SR}s. In general is the authentication in \ac{IIoT} systems very much dependent on the used protocol, as these provide optional functionalities, that can be elevated to improve security. Because of that can the attribute of the used protocol be seen in the monitoring of most of the \ac{SR}s. Though in this section, IAC is observed in a more general approach, not considering specific protocol specifications. Common protocols and their security functionalities often share similar techniques for authentication and encryption like e.g., the use of \ac{TLS} for the encryption of the network traffic \cite{Tan_2021}. In order to keep the monitorable attributes applicable to a broader scope of scenarios are attribute types generalized in this chapter. 
\begin{table}
  \includegraphics[width=\linewidth]{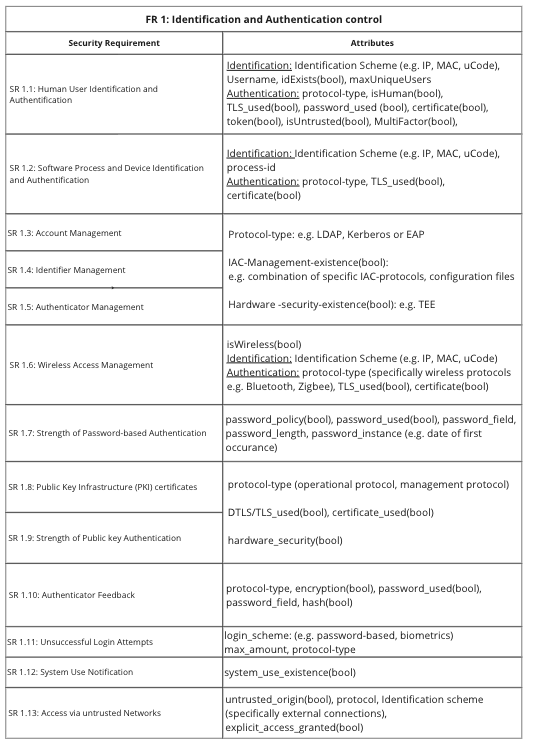}
  \caption{Collection of attributes, that can be used to monitor Security Requirement fulfillment in the FR 1 Identification and Authentication}
  \label{tab:FR1}
\end{table}

\subsubsection{IAC of users}
The identification part of the first two \ac{SR}s is described in the bottom layer of the \ac{IIoT} security architecture, the \glqq Device Layer\grqq \cite{Tan_2021}. In this layer are the different identification schemes of virtual or physical entities to the \ac{IIoT} system described. Because of that it can be derived that SR 1.1 and SR 1.2 overlap in identification attributes, as they share the possible identification schemata of the device layer. For example can the first \ac{SR} be monitored using the attribute of IP-address of the \ac{HMI} and the second requirement using the IP of the \ac{IIoT} device. In addition to the IP-address includes device identification in \ac{IIoT} a variety of schemata, which can be used as an attribute \cite{Tan_2021}. Table \ref{tab:FR1} displays these identification schemata in a single attribute. Specific examples for these could be \ac{MAC}, \ac{NetBIOS}, \ac{EPC} or \ac{Ucode} \cite{Tan_2021}. Additionally includes the second requirement, 1.2, the identification and authentication of software processes. Here can the attribute of process-ID be used for the identification of processes requiring the ID being send over the network, which depends on the process itself and the protocol used by the software for authentication. \\ Security Requirement 1.1 specifically describes the authentication and identification of humans to the \ac{IIoT} system. For that an attribute describing if the participant is human, is necessary to distinguish in between the fulfillment of SR 1.1 and 1.2. An example for this would be the identification scheme of \glqq username\grqq, as this would enable the assumption of a human process participant. The most common form of P2M-authentication in \ac{ICS} includes password-based authentication \cite{Gaiceanu_2019}. Traditional password-based authentication concludes passwords being sent in the body of the transmission protocol or in specific header fields of proprietary protocols both generating monitorable network traffic. Because of that it can be deducted, that attributes describing the existence or the location of the password string in the body or in protocol specific fields are needed. To know the specific location of such a field in the payload byte-stream or the location of the header field, \ac{SR} 1.1 also needs the protocol type as an attribute for fulfillment monitoring. Besides the described password-based approach can humans also be authenticated inherence-based with bio-metrics or possession-based with e.g., tokens. RE 2 and RE 3 of SR 1.1 also differentiate in between untrusted networks and other networks. To distinguish RE 2 and 3 is an attribute describing the trustworthiness of the network of a communication participant necessary. This attribute can e.g., be derived by assigning identifiers of entities to the documentation of the network segmentation. Another property of RE 2 and 3 is the necessity of \ac{MFA} for higher \ac{SL}s. \ac{MFA} includes the combination of two or more of the above-mentioned authentication categories knowledge, inherence, and possession \cite{Ometov_2018}. The derived attribute \glqq MultiFactor\grqq, can for example be confirmed by the detection of multiple of the in this section described authentication measures like passwords and tokens. As this includes the later described proximity-based approach it might not be possible to ensure the detection of \ac{MFA} only by monitoring the network traffic.\\ 
Security Requirement 1.2 describes requirements on the authentication of the device part of communications and software processes including \ac{M2M} and \ac{H2M} authentication. Authentication of devices in the IIoT often includes technologies like lightweight encryption algorithms over special industrial protocols, proximity-based authentication or the exchange of tokens using measures like \ac{TLS} or \ac{PKI}\cite{qureshi2018survey,Agrawal_2019}. Proximity-based authentication can be used conveniently in \ac{IIoT} networks, as it can be considered lightweight and does not rely on input \cite{qureshi2018survey}. Different methods of this technology like wire-, radio-, acoustic-, light-, gesture- or biometric-based authentication exist \cite{qureshi2018survey}. As these combine existing characteristics of the communication form, they do often not produce any network traffic itself making the monitoring of proximity-based authentication difficult. For example do radio-based schemata combine characteristics like signal strength and radio channel statistics to authenticate the device, which does not include data being exchanged over the network mediums \cite{premnath2012secret}. Possession-based schemata like the usage of a PKI include data being transferred over the medium generating monitorable traffic. Because of that is the attribute of the used protocol necessary for SR 1.2, as it includes information about the location of this authentication data in the byte-stream. Within the application layer of the network stack is \ac{TLS} used for encryption and authentication of the exchanged network traffic. The used \ac{IIoT} publish-subscribe protocols like \acs{MQTT} or request-response-based protocols like the \ac{XMPP} are dependent on \ac{TLS} \cite{Tan_2021}. As the authentication with the \ac{TLS} handshake generates network traffic, the existence of \ac{TLS} can be monitored, which allows the assumption that authentication is provided by the system. Because of that can an attribute describing the existence of \ac{TLS} authentication be added to table \ref{tab:FR1} with SR 1.2. Another form of authentication in \ac{IIoT} \ac{H2M} communications can be key-based with the implementation of PKIs exchanging public keys. This form of authentication might not be feasible in most industrial systems, as the resources of \ac{IIoT} devices are limited. Even though \ac{TLS} is the standard in \ac{H2M} authentication is an attribute for the monitoring of network traffic, generated by a key exchange, necessary for SR 1.2.\\ The SR 1.6 requires the control system to identify and authenticate all users engaging in wireless communications. In environments with \ac{SL} two or higher requires the \ac{RE} the identification and authentication process to include unique identifiers and authenticators. As described before is the IACS process very much dependent on the used protocol. To identify the network flow as being part of a wireless communication, would a monitoring system need the attribute of the protocol-type. Specific wireless protocols like Bluetooth or Zigbee allow the system to identify the flow as wireless and to set the \glqq isWireless\grqq \, attribute. After that is the compliance check similar to the previous requirements.

\subsubsection{IAC management systems}
The next three requirements include the need for different kinds of management systems. In SR 1.3 is the need for and account management solution described. In this work, an account management measure is defined as an entity or functionality in the \ac{IIoT} system, managing identified users with their identifiers, authenticators, and roles. This includes the requirements of SR 1.4 and SR 1.5, as they include the management of the specific identifiers and authenticators necessary for the authentication process described in the previous section. Such management systems can be built on commercial providers e.g., on Microsoft Active Directory, or on proprietary technologies based on e.g., the \ac{EAP} or Kerberos \cite{cisco2022}. Management systems like Active Directory are commonly used in traditional \ac{IT} enterprise systems, but as the interconnection in between the \ac{OT} and the \ac{IT}increases e.g., with the integration of cloud-based solutions into \ac{IIoT} processes, it enables possibilities to manage identifiers and authenticators on a centralized solution \cite{Wang_2019}. Monitoring the existence of such a system might be difficult. Network traffic generated by Active Directory is a combination of different authentication protocols like the \ac{LDAP} or Kerberos and often satisfies only small parts of whole authentication management functionalities \cite{microsoft21}. 

\begin{figure}[h]
    \centering
    \includegraphics[width=\linewidth]{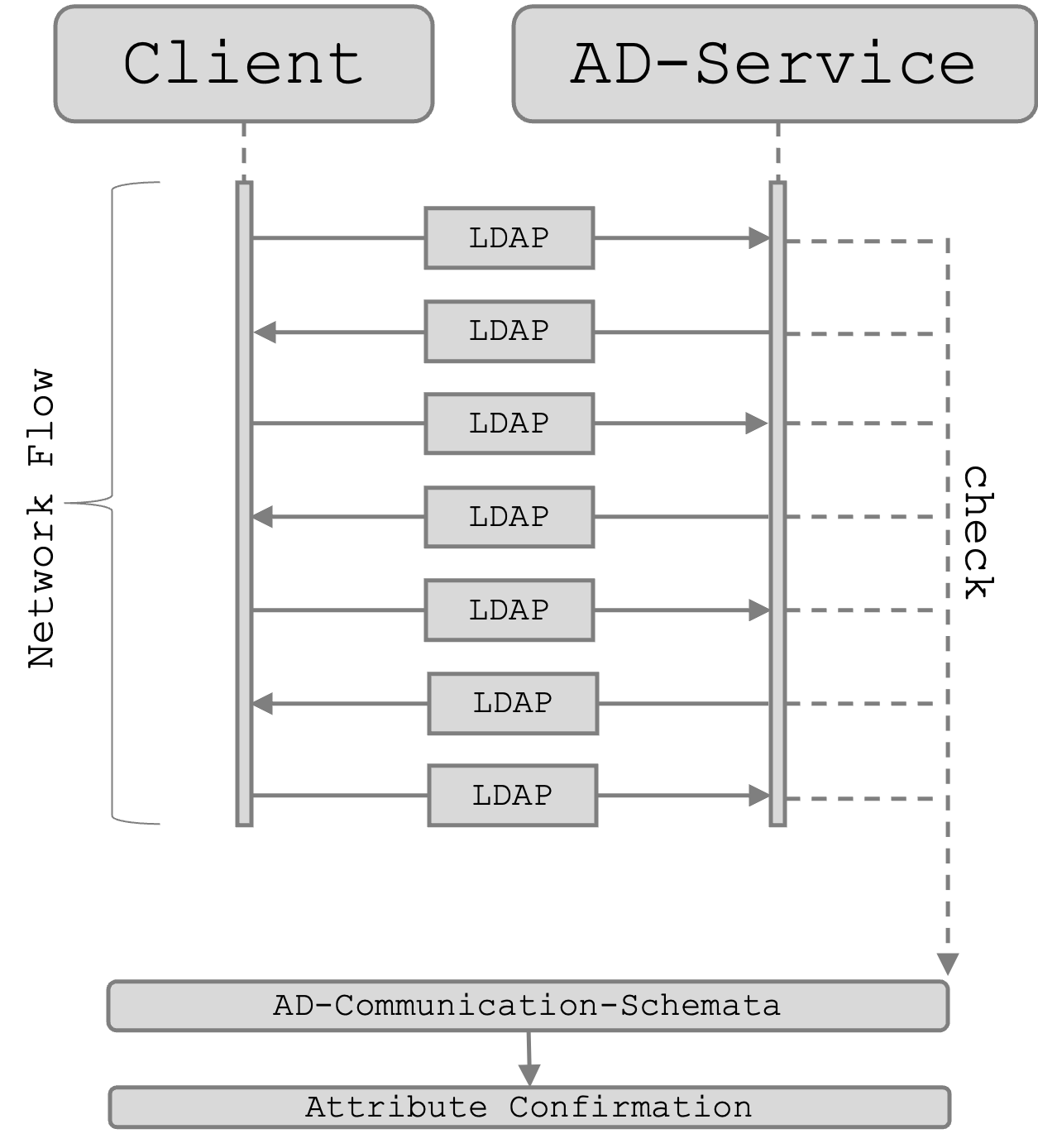}
    \caption{Attribute construction of a logical attribute for the confirmation of the Active Directory compliance status}
    \label{fig:IAC_management_figure}
\end{figure}
Furthermore does Active Directory use \ac{TLS} on all of their communications making the monitoring of packet contents difficult. Because of that might the only monitorable attribute for the SR 1.3 to SR 1.5 be the type of the protocol. By identifying specific protocols like \acs{LDAP}, Kerberos or \acs{EAP} can the \ac{CMS} assume the existence of some form of account-, identifier- or authenticator-management system. The information gained out of this attribute can further be elevated by analyzing the network flow of authentication processes and comparing it with protocol combinations of Active Directory use cases. As an example does figure \ref{fig:IAC_management_figure} show the simplified Active Directory communication flow for the deletion of a user account. This process consists of seven individual \acs{LDAP} packets being sent in between the client and the server. By counting the amount of monitored occurrences of the \acs{LDAP} protocol in a row can the \ac{CMS} set the attribute \glqq IAC-Management\grqq.

\subsubsection{Strength of password-based authentication}
Security Requirement SR 1.7 describes characteristics of the used passwords, if a password-based authentication system is deployed in the industrial network. A system according to this requirement should enforce password policies based on minimum length, character variety, and in higher \ac{SL}s restrict the lifetime of passwords. The monitoring of the requirement can be achieved by identifying passwords leading to the assumption that no password policy was enforced before. If for example two users authenticate via password and one of the passwords has more characters than the other, could the monitoring system assume that there was no function enforcing the password strength in place.

\subsubsection{Public key infrastructures}
The requirements SR 1.8 and SR 1.9 both describe \ac{IAC}requirements based on \ac{PKI}s. The first one requires the \ac{IACS} to operate the \ac{PKI}s to commonly accepted best practice, if such a \ac{PKI} is operated in the first place, or provide the capability to establish connections to existing \ac{PKI}s if necessary. This requirement can be ignored if no such system is operated and no users in the system use certification as an authorization measure. The second requirement, SR 1.9, specifies the former SR by describing requirements especially for the exchanged public keys and certificates. It includes required aspects like signature validation, certificate revocation, identification mapping, and the establishment of control of users over their private keys. The RE 1 expands the SR by requiring hardware security for the \acp{CA} and the devices controlling their private keys.\\
In general do \ac{PKI}s describe the issuing, storing, and distribution of public-key certificates in a distributed or centralized approach \cite{Maurer_1996}. The certificates consist of digital signatures based on cryptography, which are provided by a \ac{CA} and used for user authentication in every communication happening in the \ac{PKI} deploying system \cite{Krishnan_2022}. Because of the resource draining attribute of these \ac{PKI}s might the deployment in \ac{IIoT} scenarios not be feasible in some scenarios and require more domain specific approaches to \ac{PKI} \cite{Astorga_2022}. However in some scenario like e.g., smaller networks with more powerful devices, might traditional \ac{PKI}s still be viable. In combination with the improving edge capabilities and the resulting increase in processing power can more security functionalities like \ac{PKI}s be deployed in \ac{IIoT} systems \cite{Krishnan_2022}. A trade off between computational overhead, efficiency, and security can be achieved by promising \ac{PKI} variants using attribute-based encryption or infrastructures based on blockchain \cite{Krishnan_2022}. These future technologies improve specific parts like the computation of the keys, but still depend on the existence of certificates. Because of that, they are not investigated in more detail and it is focused on attributes of \ac{PKI}s in general. 
\begin{figure}
    \centering
    \includegraphics[width=\linewidth]{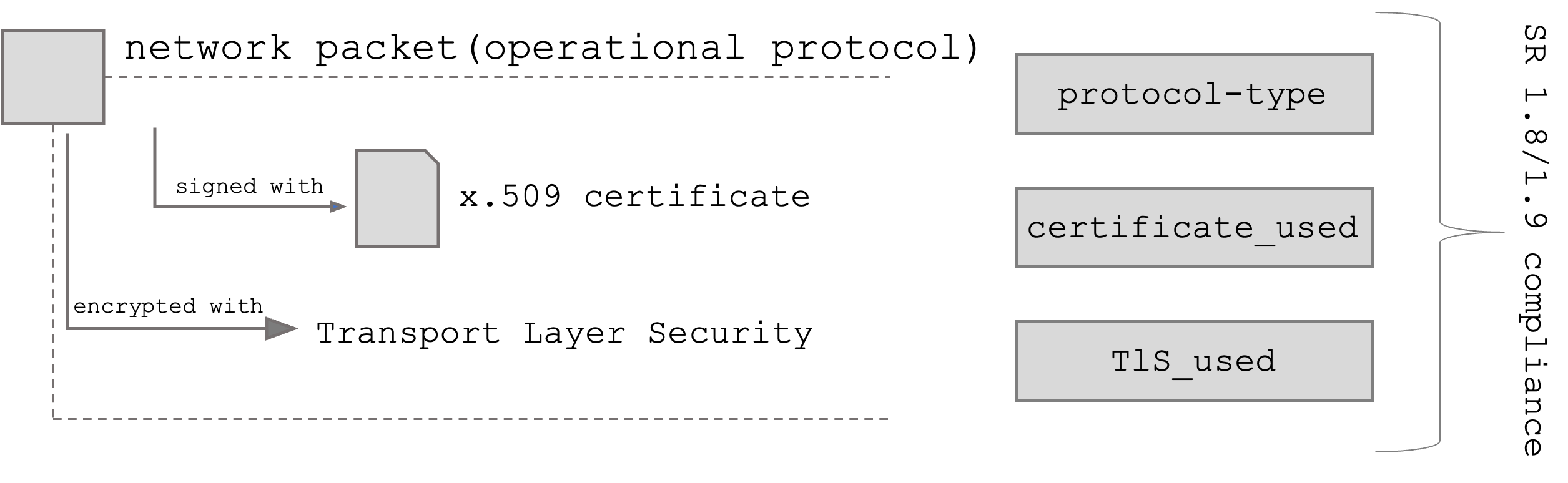}
    \caption{Example construction of an attribute extracting the compliance status to SR 1.8 and SR 1.9}
    \label{fig:PKI}
\end{figure}
In most cases \ac{PKI}s are based on x509 certificates. These are used to prove to a \ac{CA}, that the user is bound to their respective public key. The \ac{CA} signs the certificate for each user in the network, which can be done by different technical means like \ac{TLS}, the private keys of users or other ways of authentication \cite{Cooper_2008}. Each step of the \ac{PKI} life cycle is dependent on operational protocols to deliver certificates and management protocols for tasks like registration or revocation \cite{Cooper_2008}. The most popular \ac{IIoT} protocols like MQTT, XMPP, the \ac{OPCUA} protocol, and Modbus TCP support X.509 natively with the addition of \ac{TLS} or \ac{DTLS} \cite{Astorga_2022}. Because of that it can be derived, that the attribute of the protocol type is necessary for the monitoring of the SRs 1.8 and 1.9 to investigate for other attributes. Even though the mentioned \ac{IIoT} protocols do often include the capability for operational and management tasks, they are separated into two attributes. The next attribute necessary for compliance monitoring is the existence of a certificate. If the monitored traffic includes one or more certificates, it can be assumed that some sort of PKI is deployed. As the SR 1.8 specifically requires the PKI to use best practice, a combination of attributes is necessary to measure the compliance to requirement. Figure \ref{fig:PKI}\, shows the derived combination of attributes necessary to monitor the compliance of the requirements. At first is the attribute of the protocol type used to identify, if the monitored packet could be part of a PKI. If this is the case has the existence of a certificate and the according attribute to be assumed by combining different protocol specific characteristics like e.g., detecting the signature with the certificate in the payload of the packet. The last attribute consists of the best practice status of the PKI solution. To simplify the scope of this process in this work is the status of best practice defined as fulfilled if \ac{TLS} or \ac{DTLS} is used. By monitoring and combining the information of the mentioned attributes, it can be assumed that the control system is compliant to the SR 1.8 and the basic requirements of 1.9. \\ The RE 1 of the SR 1.9 requires hardware security for public-key-authentication. As the SR is not described in more detail, is the scope of the requirement to hardware security defined for users storing the private key and hardware security of the \ac{CA}. Users manipulating \ac{CA}s to provide valid certificates is a viable attack vector for attackers. Additionally can the need for hardware security of the \ac{CA} deviate from the trust problem of PKIs, as attackers perpetrating the \ac{CA}s security affects every step of the PKI life cycle \cite{Astorga_2022}. Both cases can be hindered by hardware security and should be included in the RE 1 of SR 1.9. The monitoring of such hardware extensions might be difficult. In \ac{IIoT} systems there are multiple possibilities of enabling hardware security for devices in the system like establishing Trusted Execution Environments or the use of cryptographic processors called Hardware Security Modules. Secure hardware is based on the comparison of device snapshots with verified snapshots, which are secured by measures like the formerly mentioned \cite{Lesjak2015}. As an example do some secure hardware architectures include snapshots being sent over the network to a snapshot verifier to verify the integrity of users before the control system can allow a communication \cite{Lesjak2015}. Solutions like the mentioned architecture generate network traffic, which can be monitored. This traffic can be used to define an attribute for the \ac{CMS}. The attributes definition varies based on the architecture of the hardware security measure. It could manifest as the mentioned snapshots being transferred or specific flags being set in protocols. 

\subsubsection{Authenticator Feedback and System Use Notifications}
In this section are attributes to monitor the SRs 1.10 and 1.12 derived. Security requirement 1.10 describes, that control systems in industrial networks should be capable of obscuring feedback information so that they cannot be intercepted. This includes authenticators not being sent in cleartext but e.g., with encrypted payload. It also includes measures like visual feedback of password input terminals at \ac{HMI}s. The former can be monitored as described in this chapter and the latter does not generate any network traffic, as it is part of the design and development of the \ac{ICS} itself. Measures of authentication in \ac{IIoT} systems were described in section \ref{FR:1} and involve authenticators being sent over the network to authenticate or exchange keys. To satisfy the SR 1.10 none of the actual authenticators should be monitorable by themselves. Authentication processes using passwords should for example either encrypt the whole communication or send the passwords as its hash representation. For the monitoring design, should a \ac{CMS} have access to the password-based attributes described before including the location of the password in the network packet. This ensures that passwords are not being sent in cleartext. Additionally does the system need an attribute representing the encryption status of the network flow. This value could either be set by detecting the existence of \ac{TLS}/\ac{DTLS} or with the attribute of the protocol type. \\ Security requirement SR 1.12 requires \ac{IACS} to display notification messages before the authentication process can start. This requirement is directed to \ac{H2M} authentication, as it relies on describing visual notifications for the operating systems of \ac{HMI}s over the use of logging-based notifications in deeper network layers. As measures implementing this requirement do not produce any network traffic, or the existence of such measures cannot be inferred by combining network level attributes, only one attribute is derived. This attribute includes a boolean value describing the existence of such a notification system. 

\subsubsection{Unsuccessful login attempts and untrusted networks}
According to the SR 1.11 should the control system enforce a maximum of allowed unsuccessful login attempts. Before the monitoring system can count unsuccessful login attempts is an attribute describing the adherent login scheme necessary. \\
In addition to the login scheme needs the protocol type to be set accordingly. The protocol type specifies the used login schema, as the monitoring system needs the attribute for the location of the transferred authenticators. The monitored network packets for this requirement could be the actual authenticators being transferred or authentication management packets being sent by the control system after a failed login attempt. These packets would include log messages stating that the login attempt was unsuccessful and differ in protocol type, as they could also consist of network management protocol types like the ICMP. A \ac{CMS} should alert on cases where more attempts are being monitored than allowed by the control system, as this would allow the assumption that the control system lacks the functionality to limit these attempts. Because of that an attribute describing the maximum allowed attempts must be present for the monitoring of this requirement. \\ The last requirement of the FR 1 describes the handling of access to the system via untrusted networks. Control systems should provide the capability to monitor and control such communications and demand explicit access approval on higher \ac{SL}s. At first should the understanding of untrusted networks in the scope of this work be defined. We define an untrusted network in this context, as a connection coming from an external IP-address outside the scope of the \ac{ICS} or a manually assigned network zone. This could be a communication attempt over the internet or a connection attempt within a virtual private network of a different network zone with a different control system and lower \ac{SL}. The first required attribute describes if the communication has an untrusted origin in the first place, as this is necessary to further specify the compliance monitoring in the scope of the \ac{RE}. Setting this attribute during the monitoring process requires the attributes of the protocol-type and the source-IP. A suspicious combination of a protocol not having the possibility for \ac{IAC} like e.g., the HTTP protocol in combination with an IP-address hinting to an external connection like e.g., the IP of the internet gateway, would allow the monitoring system to assume that there was an access granted from an untrusted network.

\subsection{FR 2: Use Control}
\label{sec:FR2}
The second \ac{FR} defines requirements for the control system, that come in place directly after a user was authenticated \cite{Kobes2021}. It describes how the assignment of privileges has to be controlled before they actually are granted with the goal to restrict allowed actions for authorized users and minimized potential privilege misuse. As the measures for Use Control are closely related to the measures of the first FR are the monitorable attributes somewhat similar.
\begin{figure}[h]
    \centering
    \includegraphics[width=\linewidth]{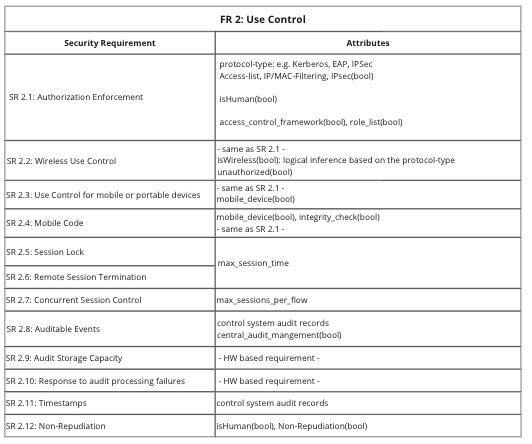}
    \caption{Collection of attributes, that can be used to monitor Security Requirement fulfillment in the FR 2 Use Control}
    \label{fig:FR2_UC}
\end{figure}
\\ The first \ac{SR} describes, that the control system should be able to enforce the authorization provided by the previously happened identification and authentication process \cite{Kobes2021}. In general is authorization closely linked to access control and describes the necessity for the prevention of the unauthorized use of \ac{IIoT} resources \cite{Tan_2021}. As described in the previous sections is authentication mostly based on authentication schemes like Kerberos, PKIs or other cryptography based measures, which all lack functionality for authorization and access control or do only provide very limited variations \cite{Tan_2021}. Advanced authorization frameworks adapted to the \ac{IIoT} like role- (RBAC), attribute- (ABAC), or identity-based access control (IBAC) are still part of the ongoing research as they are confronted with \ac{IIoT} specific challenges \cite{Salonikias_2019}. So does RBAC and ABAC require the exchange of a huge amount of attributes or role definitions in between the access control components, which has to be implemented in an efficient and secure matter \cite{Salonikias_2019}. As these advanced authorization measures are most likely not implemented in current \ac{IIoT} systems in practice do simpler measures form the current state-of-the-art. These include authorization based on access control lists, MAC- or IP- address filtering, IPSec policies or the mentioned authentication protocols with their limited functionalities \cite{Tan_2021}. \\ In table \ref{fig:FR2_UC} are the attributes, that can be used to monitor the compliance to the second FR, displayed. As the general requirement includes only the enforcement of authorization for human users, it is necessary to distinctly identify human user identifiers by a separate logical or manual attribute. In higher \ac{SL}s does the first \ac{RE} include the authorization for all users including humans and devices. Monitoring attributes or techniques in that context could include the checking for access lists being transferred, IP/MAC-filtering configurations or monitoring for the existence of IPSec. These functionalities provide basic authorization capabilities, which is enough for the lower \ac{SL}s and the inferred general authorization requirement. In addition there are the remaining three \ac{RE}s including permission mapping to roles, a functionality for supervisory override, and a dual approval approach for access granting \cite{Kobes2021}. As these are part of a multi-component access control framework as described above, they depend on the individual implementation such an approach. To include them in this work, a manual attribute listed in table \ref{fig:FR2_UC}, representing the requirement for access control frameworks, is necessary.\\ The next three requirements SR 2.2, SR 2.3 and SR 2.4 expand the previously mentioned attributes for authorization on wireless connection and on portable devices in general. In the basic context of SR 2.2 can the mentioned attributes of SR 2.1 be expanded with the factor of monitoring for wireless protocols like Bluetooth and further specify if such protocols have built-in authorization functionalities. In higher \ac{SL}s it is necessary that the wireless user is authorized, which could for example be checked with a logical attribute derived from a list of expected wireless communications. For SR 2.3 and SR 2.4 is an attribute necessary to identify if a user is represented by a mobile device. Especially in the context of SR 2.4 should a control system should be capable to restrict the usage of mobile code and further check the integrity of executed mobile code. \\ The next few requirements concern specific requirements on sessions. As the concept of a session is not further described in the IEC 62443, it is understood as an overall task in between two entities within the process including multiple connections with the same task in mind. Attributes describing different properties within the session context, that can be used to monitor the listed requirements, can be defined. So can the requirement for a capability of the \ac{IACS} to lock or terminate sessions or more specific remote session be monitored with an attribute defining the maximum allowed time for the whole session. If such a threshold is met it can be inferred logically that the session is not terminated correctly.\\ After that do SR 2.8, 2.9, SR 2.10 and SR 2.11 include requirements in the context of audit events. This includes the requirement for the existence of a central audit management system and the the requirement for actual hardware-based properties like storage capacity. As can be seen in table \ref{fig:FR2_UC} are some of the more specific audit regarding requirements based on hardware properties and can hardly be monitored. In case of SR 2.11 describes the requirement, that audit events should include timestamps. To check the compliance to this requirement does the monitoring system need access to a audit record to then check such a property.\\ The last requirement of FR2 includes non-repudiation as a requirement for \ac{IACS} systems in targeted SLs of three or higher. This requirement is one of the commonly known \ac{SR}s for the enterprise domain. As can be seen by the fact that it only represented by a single \ac{SR} and in addition only for higher \ac{SL}s, it shows that it is placed in a sub-ordinary role within the \ac{IIoT} domain. It is defined as a property, which infers that an author of a message or the originator of an action cannot deny the action or the sending of the message \cite{Tange_2020}. To represent this part within the monitorable attributes is a logical or manual attribute required in combination with an attribute describing if the user is identified as human.

\subsection{FR 3: System Integrity}
\label{sec:FR3}
In general does integrity as a security goal concern the trustworthiness, consistency, and accuracy of data over its entire life cycle, which includes the occurrences of data-in-motion and data-at-rest \cite{Tan_2021}. This trustworthiness can be seen as violated if the modification or manipulation is detected by security functionalities \cite{Tan_2021}. The FR 3 expands system integrity on this definition including the integrity of physical assets in operation and non-operational phases. Table \ref{tab:FR3} shows a collection of attributes that can be used to measure compliance with the \ac{SR}s of the third \ac{FR}. These are derived within the same procedure used in the previous section. In this case are possible implementations of integrity ensuring measures observed to find attributes and characteristics, that can be used to assume the compliance status. This table is not exclusive, which means that additional attributes might appear in specific scenarios.
\begin{table}[h]
  \includegraphics[width=\linewidth]{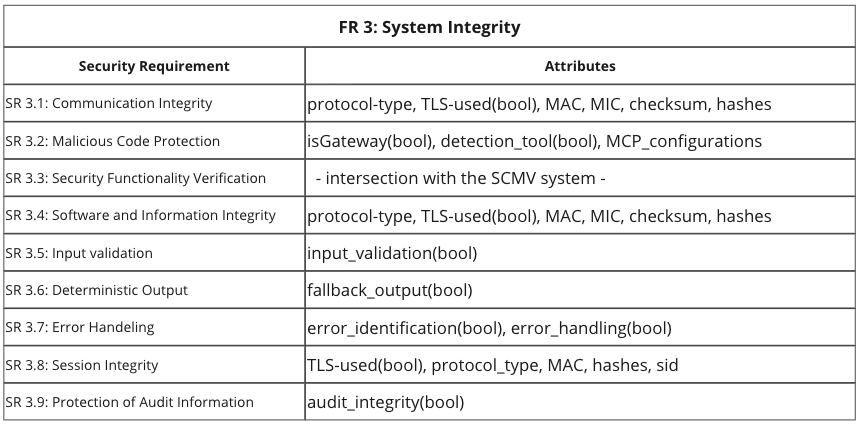}
  \caption{Collection of attributes, that can be used to monitor Security Requirement fulfillment of the FR 3 Integrity}
  \label{tab:FR3}
\end{table}

\subsubsection{Communication, software, and information integrity}
Integrity measures for logical assets, including data, in \ac{IIoT} are closely related to the authentication attributes presented in section \ref{sec:FR1} and \ref{sec:FR2} and depend on the specific protocols in use. Based on the same argumentation is integrity in \ac{IIoT} described in a general way to create a protocol independent view on the monitorable attributes, though providing protocol specific examples. \\ The first requirement is SR 3.1, which requires the protection of all communications in the network. It also includes the \ac{RE} 1, which further specifies the SR to use cryptographic integrity protection schemes in \ac{SL}s three or four. As communications include packets being sent over the network, they consist mostly of data-in-motion. In the scope of this work is data-in-motion defined as transmitted data being part of communications over the different network mediums and layers, which allows us to handle it completely in this section. The cryptographic protection of the integrity of communications in \ac{IIoT} is heavily reliant on IPSec and \ac{TLS} or \ac{SSL} mechanisms \cite{Tan_2021}. As the existence of \ac{TLS} can easily be monitored via the network traffic, an attribute stating the existence of \ac{TLS} is necessary for the monitoring of the integrity. Monitoring attributes allowing the assumption, that \ac{TLS}/\ac{DTLS} is used in the communication is enough to fulfil the requirement SR 3.1 with RE 1. As mentioned before might \ac{TLS} not be feasible in every scenario, as of the resource constraint of \ac{IIoT} nodes. Encrypting the whole payload via a PKI or 256-bit symmetric AES keys can produce too much overhead. Because of that are other integrity protection schemes often used in \ac{IIoT} systems, although popular application layer protocols like e.g., \acs{MQTT} or the \ac{CoAP}, would support \ac{TLS} or \ac{SSL} \cite{Tan_2021}. Digital signatures and the use of hashing algorithms are the most prominent alternatives to the use of \ac{TLS} \cite{Esposito_2018}. As they do not rely on the computational power of \ac{IIoT} nodes to encrypt whole network packages, there usage is more viable in a broader spectrum of \ac{IIoT} systems. Comparable to the problems mentioned in the previous sections do digital signatures for integrity purposes suffer under the same problems as signatures for authentication purposes, which make them challenging to deploy in every \ac{IIoT} scenario. As these challenges primarily rely on the use of more lightweight encryption schemes like e.g., elliptic curve cryptography or more lightweight PKI solutions like e.g., group-based certificates, they do not impact the monitoring of the FR by itself. The existence of a certificate, for example a x.509-certificate in the \acs{MQTT} protocol, can be seen as a monitorable attribute for compliance monitoring \cite{Esposito_2018}.

\subsubsection{Malicious code protection and security functionality verification}
The requirement SR 3.2 states that the capability to prevent, detect, report, and mitigate on effects of malicious code should be given by the control system. This includes malicious code in a traditional sense, but also includes other forms of unauthorized software. The \ac{RE}s of the \ac{SR} further specify the placement of the protection mechanism. So does the lowest \ac{SL} only require a protection in general on any place in the system, the first enhancement specifically the protection of entry and exit points of the system, and then in addition to that does the second enhancement require the central management of protection measures. Checking if such a protection measure is implemented can be complicated. In general does malicious code include the definition of malware, which is the most prominent form of malicious code \cite{Shah_2020}. Various evolving forms of malware try to invade \ac{IIoT} systems by utilizing vulnerabilities in applications, operating systems, network devices, and databases \cite{Sun_2021}. To protect against malware are software tools used to detect the malicious code, prevent its spreading, and then finally delete it. Similar to IDS categories there are static, dynamic, and hybrid malware analysis technologies \cite{Sun_2021}. Static analysis tool analyze executable code within a system, by decompiling and disassembling the code looking for matching signatures among known malware \cite{Sun_2021}. In comparison to that do dynamic malware analysis technologies not observe the code of the malware, but include the behavior of the program and its impact on the system \cite{Sun_2021}. The combination of both of the technologies is combined among the class of hybrid malware analysis technologies. Despite the analysis method and deployment location are these technologies integrated into malware detection tools. To check for the compliance to the malicious code related SR 3.2 must an attribute reflecting the existence of such tools be monitored. In addition to the existence of such tools as they are, is the location of the deployment within the network of great relevance. That includes an attribute stating if the component of the implemented detection tool is impersonating a gateway role to an external system. The second enhancement requires a central instance to manage the malicious code protection in the system. By checking a central management instance for malicious code logs or configurations managing the deployment of the analysis tools can the compliance be checked in addition to the previously mentioned attributes.\\
In general do security functionalities or security solutions exist within the \ac{IIoT} system to address specific threat vulnerabilities with the goal to reach a specific \ac{SL} \cite{Khan_2018}. This intercepts with the definition of \ac{SLM}s given by the IEC 62443. To be compliant with SR 3.3 should an \ac{IACS} provide the capability to verify, that these functionalities function as intended. In the basic representation of the requirement this includes only the verification during \ac{FAT/SAT} within the commissioning phase of the \ac{IACS} life cycle. The first \ac{RE} is relevant in systems with \ac{SL}s three or four and requires the security functionality verification to be automatic, but still only during the commissioning phase. In addition to that in systems of \ac{SL} four must the verification be given during the normal operation phase of the \ac{IIoT} process. Automatically verifying the functionality of implemented security functionalities is in parts fulfilled by the \ac{CMS} system itself, which verifies the compliance the IEC 62443 security measures, including the security functionalities, within the monitoring layer. The description of the SR 3.3 is a contradiction to the foundational livelihood of the \ac{CMS} system including the definition of monitorable attributes. To verify the compliance with SR 3.3, would the \ac{CMS} system have to check itself, if the verification of functionalities is going according to the requirement. Because of that this can the requirement not be monitored, as the \ac{CMS} system fulfills the general requirement by itself. Even though there could some special cases be, where verification logs or other configurations can lead to assumption that another functionality verification system is in place, it is excluded in this section of the list with monitorable requirements.

\subsubsection{Input and Output Integrity}
The SR 3.5 requires the control system to validate the syntax and content of all inputs, that directly affect the control system itself. Such inputs could for example be process deviating and control inputs changing production or security related properties. Checking the existence of  measures implementing this requirement might be complicated, as such validation features are incorporated into the \ac{IACS} software. Because of that there might not be a monitorable attribute and for the verification of the requirement has the according attribute to be set manually. This can happen during the \ac{FAT/SAT} phase by manually provoking declined inputs. In the same context can the compliance to SR 3.6, requiring the control system to incorporate deterministic output values, not be checked. Concrete does the SR include components and control instances to fall back to a deterministic value in case of errors or abnormal behavior. The according attribute can also only be set manually during the commissioning phase, which includes the necessity for a boolean attribute in table \ref{tab:FR3}.

\subsubsection{Error, Session, and Audit Integrity}
The last three \ac{SR}s of the third \ac{FR} include measures describing the integer handling of errors, sessions, and the protection of audit information. In regard to the error handling, should the \ac{IACS} be able to identify and handle them in a way enabling effective remediation. This also includes, that the error messages should not provide information about the \ac{IACS}, that can be used by potential attackers. The second part of the requirement is very much subjective and cannot be monitored, as the border in between necessary and vulnerable information depends on the system. In case of the identification of error messages can the checking be possible by monitoring for special network management protocols including error codes. Even in systems where network traffic is generated based on the management protocols might the compliance check include the manual setting of an attribute. The monitoring of the \ac{IACS}, identifying and handling an error includes the simulation of an error scenario, which can usually be done within the \ac{FAT/SAT} of the commissioning phase. Individual monitoring solutions include specific monitoring attributes depending on the \ac{IACS} configuration. Overall is the necessity of two boolean attributes documented rating the compliance to error identification and handling.\\ The \ac{SR} 3.8 describes, that the \ac{IACS} should be capable of ensuring the integrity of session and identifying invalid session IDs. Integrity protection for sessions depends on the used application layer protocols. Most of the common application layer protocols like \ac{HTTP} or \ac{MQTT} implement security functionalities including integrity with the use of \acs{TLS}/\acs{SSL}. By detecting the existence of \acs{SSL} can the compliance checker confirm, that the \ac{IACS} uses session integrity. Checking the second part of the requirement, which includes the \ac{IACS} identifying and blocking invalid serial IDs requires the monitoring systems to compare protocol header information to derive the session properties. These properties can then be used as additional information input to determine if a session ID is invalid. The detection of an invalid session ID can then lead to the assumption, that the \ac{IACS} might not be able to detect and to end invalid sessions. \\ The last \ac{SR} SR 3.9 describes the \ac{IACS} requirement for the protection of audit information and tools from unauthorized access, modification, and deletion. The protection of audit information depends on the storage context of the audit logs. For example can the audit logs be stored in a network attached storage, that is protected by a restrictive firewall, which is the best case scenario or the audit logs are collected within each of the network components itself. The simple proving of the existence of audit logs in general might not be sufficient to allow the assumption, that the SR is fulfilled. Because of that has the boolean attribute for the compliance measurement to be set manually.

\subsection{FR 4: Data Confidentiality}
The fourth \ac{FR} defines requirements to the \ac{IACS} in regard to the security objective confidentiality especially for data in-motion and in-rest. It is not described, that every information transferred has to be encrypted, but the \ac{IACS} should provide the capability in cases where confidential information is exchanged. Additionally must in these situations the chosen confidentiality measure fulfill special requirements. An example collection of monitorable attributes is provided in table \ref{tab:FR4_confidentiality}. The security objective confidentiality within the \ac{IIoT} domain can be defined as the protection of the information flow in general \cite{Chhetri_2017}. This general definition includes aspects like data encryption, access control, network segmentation, and privacy \cite{Tange_2020}. As some of these aspects are handled within other \ac{FR}s do the monitorable attributes of this FR mainly focus on data encryption as the main sources of security. The main example for that is access control, which was already discussed in section \ref{sec:FR1}. Within the scope of that definition are monitorable attributes for the confidentiality based \ac{SR}s described in the following. \\
The first requirement SR 4.1 describes requirements in regard to confidentiality of information, that explicitly require read authorization. Because of that optionality must the cases where confidentiality protection is required be identified with a boolean-based attribute. A concrete initialization of the attribute can for example be achieved by comparing the IP addresses within the monitored communications with a list of confidential zones. In zones with higher \ac{SL-T} is the general definition of the \ac{SR} is specified further by two \ac{RE}s. The first \ac{RE} adds the mandatory protection of confidentiality in cases where information is transferred via or stored in untrusted networks. A specific definition of untrusted networks in this context is not given within the IEC 62443. In this work are untrusted networks defined in the context of confidentiality as network zones being external to the \ac{IIoT} system itself or being internal in another zone, that has a lower security target defined. To monitor compliance for this special case is a boolean attribute defining if a communication is affected by the provided definition necessary. Based on the identifiers either hinting to external networks or by comparison with the security targets of the network zones can the compliance monitoring solution set the value of that boolean value accordingly. The last \ac{RE} restricts this further by expanding the required confidentiality on the whole network including all zones and information leaving the network boundaries. Implementing information confidentiality in general is achieved by protocol specific network encryption algorithms. For example are in the IIoT domain application layer protocols like \acs{HTTP} or \acs{MQTT} the most prominent, which implement confidentiality with \ac{TLS}. An example for a different layer is the usage of security functionalities of Hadoop \acp{RPC}s within the data processing layer \cite{Tan_2021}. In general can it be stated that most \ac{IIoT} systems rely on \acs{TLS}/\acs{SSL} or IPsec as their preferred security confidentiality measure \cite{Tan_2021}. Because of that are \acs{TLS} and IPsec both displayed as monitorable attributes within table \ref{tab:FR4_confidentiality}, noted that these can be seen as examples for other measures that represent monitorable network security. Proving the existence of such measures only by deploying passive monitoring is easily achievable as most network monitoring tools like Suricata \footnote{https://docs.suricata.io/} or Snort \footnote{https://docs.snort.org} include \ac{TLS} specific pre-processors or logging capabilities. 
\begin{table}[h]
  \includegraphics[width=\linewidth]{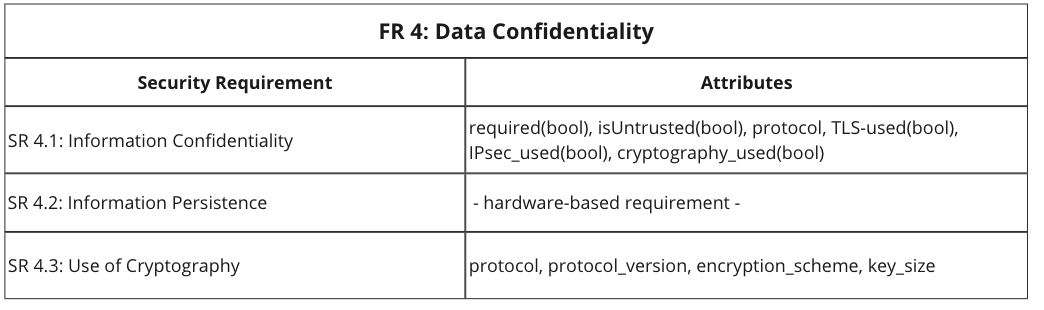}
  \caption{Collection of attributes, that can be used to monitor Security Requirement fulfillment of the FR 4 Confidentiality}
  \label{tab:FR4_confidentiality}
\end{table}
Additionally can the first requirement be linked to the third by including an attribute defining if the confidentiality measure uses cryptography. The SR 4.3 requires \ac{IACS} to use commonly accepted security industry practices, if cryptography is used. Measuring the compliance to this requirement presumes the existence of used cryptography. Because of that argumentation, there are two scenarios where this requirement becomes relevant. The first is the case when a protocol implements the security functionalities as part of the protocol functionalities e.g, by using \ac{TLS} or IPsec. In such cases can the protocol version be used to rate the security of the cryptographic solution. This attribute is widely used in practice to monitor standard compliance, as presented in the results of the compliance monitoring market study. If confidentiality is based on protocol independent encryption, for example when a industrial protocol does not support encryption natively, can the security rating be achieved by identifying the encryption scheme with the size of the used encryption keys.\\ Finally comes the missing \ac{SR} of the confidentiality \ac{FR}, SR 4.2, which describes the necessity for an \ac{IACS} security measures to purge confidential information from components being de-commissioned or leaving the system in other ways. This aspect is solely hardware-based and cannot be monitored by active or passive monitoring solutions. Because of that has the requirement compliance to be set manually.

\subsection{FR 5: Restricted Data Flow}
The fifth \ac{FR} is a direct result of the zones and conduits concept of the IEC 62443 standard transferring the theoretical concept into a realizable guideline. Overall do the \ac{SR}s described here require asset owners to determine the information flow within and across the zones and configure the conduits delivering the information. These conduits can be seen as the communication interfaces in between different zones. As these zones have different \acp{SL-T}, they should be segmented logically or based on hardware. By deliberately defining the conduits in between these zones, they can be configured to achieve compliance with the requirements of the fifth \ac{FR}. Table \ref{tab:FR5} displays an excerpt of monitorable attributes for the SRs 5.1 up to SR 5.4. The first \ac{SR} describes the general requirement for the \ac{IACS} to be capable to logically segment control-, non-control, and critical system networks. In \ac{IACS} with a \ac{SL-T} of two or higher does the first RE require the network segmentation to be physical. The second \ac{RE} further expands the requirement layers with the necessity for non-control networks to be independent from control networks. This means that the non-control network can work independently without being dependant on control directions over network conduits. On the highest \ac{SL} specifies the third \ac{RE}, that critical networks must be isolated logically and physically meaning that no conduits are configured to these systems.\\
In general does network segmentation describe the partitioning of the network into smaller segments based on similarity factors like criticality or management authority \cite{Stouffer_2015}. In more detail can segmentation be divided into logical and physical segmentation and include implementations based on each of the \ac{IACS} layers \cite{Sanchez_2020}. Logical segmentation can be based on encryption e.g., virtual private networks, or by implementing different network filtering technologies, whereas physical segmentation includes the segregated segment using their own network infrastructure and only being connected to the others via the configured network conduits \cite{Stouffer_2015}. An example for the monitoring of logical segmentation could be based on IP sub-netting or by actively detecting firewall configurations within components. Checking the requirements of physical network segmentation might not be possible, as this includes hardware aspects like separate networks. Even if the \ac{CMS} system can monitor the traffic within the physically separated network, it cannot prove the hardware based segregation. Because of that can a boolean attribute be set manually during the commission phase. Compliance to the second \ac{RE} requiring non-control systems to be independent includes, that there is no process relevant communication in between the non-control system and the \ac{IACS}. Such a process relevant communication could include the control system being used as an authentication authority or missing conduits to segments interacting with the examined segment. Another case of missing compliance to the requirement of independence would be the usage of ICMP messages from the control system being relevant within the process. Because of that can the attribute of protocol type be used to create monitoring rules for the network segmentation. The described independence has to be represented by an attribute, which is logically derived from the gathered evidence. An example for this would a monitored communication be in between the control system and a relevant system, which is required to be independent. The communication includes a mandatory control process step based on the ICMP protocol necessary to continue the process. Such a mandatory communication would be the control system being used as a traffic conduit inferring dependence. By combining monitorable traffic attributes can the independence be derived to a logical boolean attribute.
\begin{table}[h]
  \includegraphics[width=\linewidth]{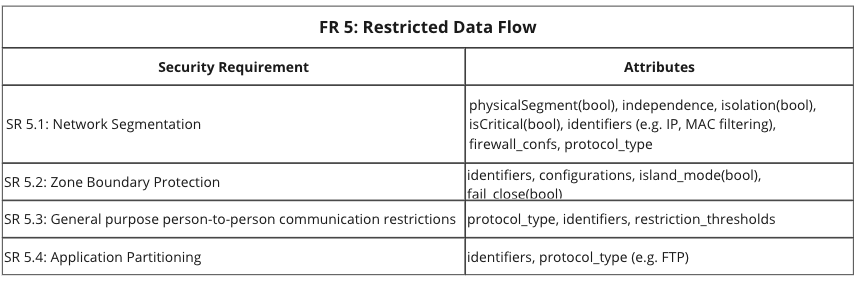}
  \caption{Collection of attributes, that can be used to monitor Security Requirement fulfillment of the FR 5 Network Segmentation}
  \label{tab:FR5}
\end{table}
The last \ac{RE} requiring logical and physical isolation cannot be checked as a whole, as the physical isolation and segmentation cannot be monitored as described. Logical isolation can only be checked for compliance if the \ac{CMS} system is implemented in its own representation within the critical network. Then the system can monitor for communications targeting external entities, which would allow the assumption that the system is not isolated.\\ The next SR 5.2 requires the control system to provide the capability to monitor and control communications at zone boundaries with the intend to enforce the zones and conduits model. In the field of \ac{ICS} and \ac{IACS} are \acp{BPD} used to achieve the protection of the network boundaries by enforcing security related policies \cite{Stouffer_2015}. There are multiple measures functioning as \ac{BPD}s possible, like network routing hardware, network filtering or software-based measures controlling the network flow \cite{Stouffer_2015}. Making an assumption about the existence of such devices and the according security policies might be possible by looking for security configurations. These could be in form of configuration files within the actual devices or in form of network policies being distributed by the control system. The first \ac{RE} specifies the requirement further by requiring the adoption of the least privilege concept. This includes communications leaving the boundaries to be denied by default and only be accepted by exception. By detecting communications across the network boundaries, which are not defined within the known behavior, can the assumption be made, that whitelisting is not implemented on the boundaries. The next \ac{RE} requires the \ac{IACS} to be possible to isolate the control network from the other segments, if a zone is compromised. A compliance check to this requirement is based on network architecture, which cannot be monitored that easily. Because of that is an attribute, either manual or logical, necessary to track this compliance factor. The third \ac{RE} states that the \ac{IACS} should be capable to close the system boundary in a fail state. This would deny the spread of security incidents through the network. It can only be checked with a manual attribute during the commission phase or by looking for such a configuration within the configuration files. Actively scanning the configuration files requires the boundary protection measure to be well known.\\ The SR 5.3 requires the \ac{IACS} to restrict general \ac{P2P} communications and expands this requirement to forbid them on targeted \ac{SL}s three or higher. The monitoring for prohibited \ac{P2P} can be achieved by defining \ac{P2P} protocols e.g., \acs{HTTP}, and combine that information with the identifiers of the communication partners. If such a communication is detected in between two human process participants can the monitoring system assume that the \ac{IACS} does not prohibit them. In case of only restricting the \ac{P2P} communications it depends on the actual restriction measures. If for example the restriction is based on bandwidth limitations, than a threshold can be defined as a traffic attribute, which can then be used to continuously monitor the enforcement of this restriction. On the other hand if the restriction is based on physical isolation it would not be possible to check the compliance. \\ The last SR within FR 5 requires the partitioning of data, applications, and services in accordance to criticality of the network zones. A full compliance of the requirement might not be possible as it is described only vaguely and depends on the actual visibility of the services of each device within the network. As an example is the protocol type shown, as it can be used to detect file transfer across the network boundaries, which would be a contradiction to the data partitioning.\\ In summary it can be stated that the fifth \ac{FR} is very dependant on configuration files of boundary protection measures and includes hardware-based requirements. This allows the conclusion, that the monitoring of these requirements is very much dependant on active components analyzing the configuration and manual compliance checks. 

\subsection{FR 6: Timely Response to Events}
The sixth \ac{FR} highlights the necessity for a control system to establish security policies and procedures to respond to security violations in an appropriate time. This also includes the definition of lines of communications and mechanisms to collect and correlate forensic evidence. Overall does this section include two separate \ac{SR}s. The first states that audit logs should be accessible all the time either in an arbitrary form or in \ac{SL}s three or higher with programmatic access measures. An audit log can be defined as a record of security related events of a predefined type at a specific timestamp \cite{Leander_2019}. Accordingly do these logs contain valuable information about the security functionalities of the system and must be protected in rest and in motion \cite{Leander_2019}. Even though this protection must be implemented, it is necessary to enable an efficient access for eligible entities as described by the first \ac{SR} of the sixth FR. The second requirement states that there should a measure be implemented, that continuously monitors the network for security violations without affecting the operational performance of the control system. In this context it should be noted that the continuous monitoring of the \ac{SR} is not fulfilled by the compliance monitoring solution of the monitoring layer itself. This is the case as the compliance monitoring only monitors for abnormal behavior in the context of compliance for specific requirements and does not include other forms of abnormal behavior e.g., violations caused by an attacker.\\
Both of the described requirements can either be implemented by an individual solution e.g., storing event logs generated by an IDS, or by a \ac{SIEM} system, which is responsible for the collection and the correlation of the security events building the audit logs \cite{Bhatt_2014}. Therefore do such \ac{SIEM} systems manage the accessibility aligned with the first requirement.\\ Table \ref{tab:FR6} shows the excerpt of monitorable attributes for the sixth \ac{FR}. In general is a logical attribute stating the existence of the audit logs itself necessary to check parts of the fulfillment of both requirements. In case of the first requirement could an attribute stating if a \ac{SIEM} is deployed also be used to check the fulfillment status. Such an attribute could for example be set manually within the commission phase.
\begin{table}[h]
  \includegraphics[width=\linewidth]{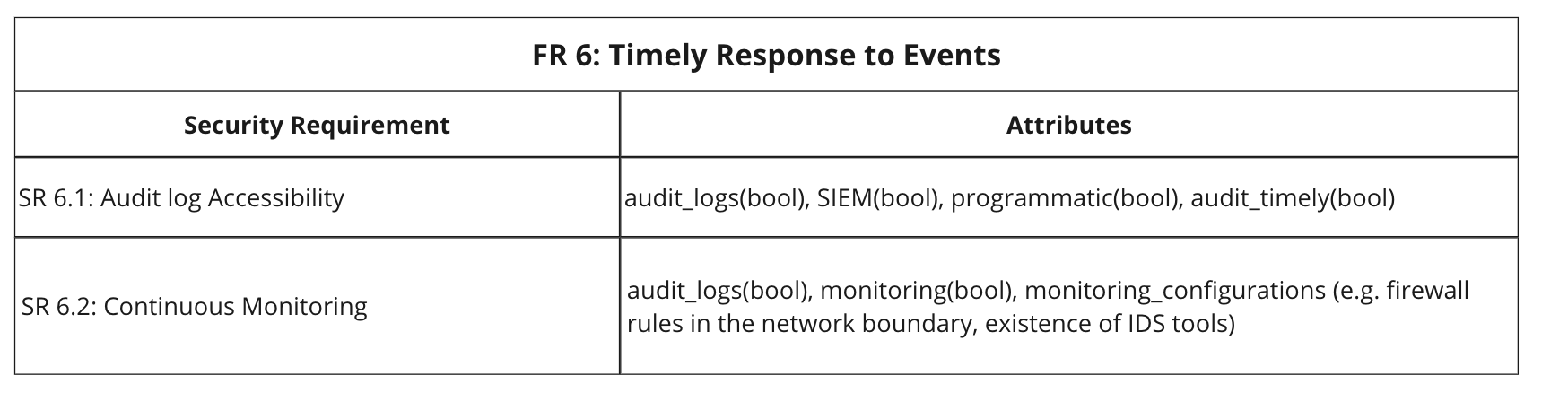}
  \caption{Collection of attributes, that can be used to monitor Security Requirement fulfillment of the FR 6 Timely Response to Events}
  \label{tab:FR6}
\end{table}
\\ For the second requirement varies the existence of a security measure implementing the continuous monitoring on the network zones. A monitoring concept could include the deployment of different types of IDS as described in the background chapter or could just include different types of firewalls. Because of that it is difficult to ensure the full compliance to the requirement based just on network related attributes. A logical attribute is necessary to verify if the continuous monitoring is implemented in the network. It could for example be inferred by discovering IDS tools within the network or by looking for firewall configurations in all of the network boundaries.

\subsection{FR 7: Resource Availability}
The last \ac{FR} describes \ac{SR}s in the context of availability. In general, it can be divided into three targeted principles an \ac{IACS} should implement. The first requires the system to be protected against various types of deny-of-service events, the second to achieve a system resilience if system functionalities are unavailable, and the third describes that security incidents should not affect essential system functionalities. These are depicted by eight different \ac{SR}s, which are presented in the following.
\begin{table}[h]
  \includegraphics[width=\linewidth]{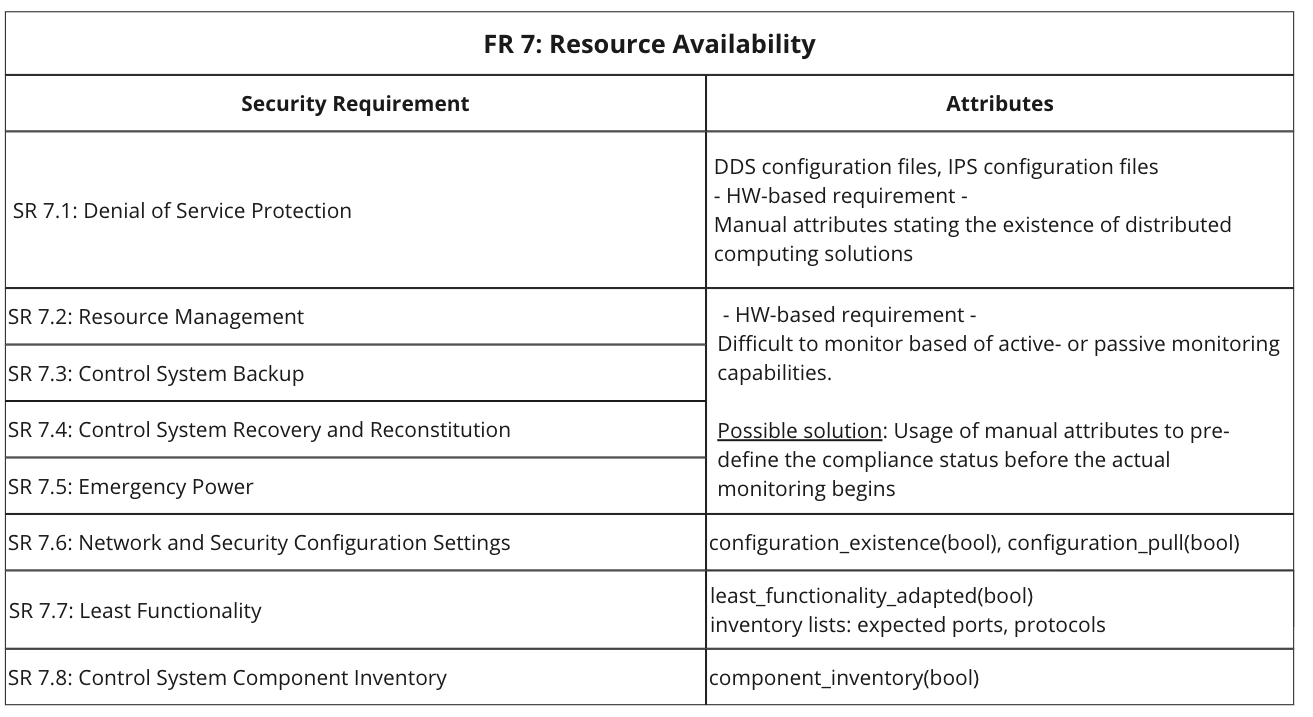}
  \caption{Collection of attributes, that can be used to monitor Security Requirement fulfillment of the FR 7 Availability}
  \label{tab:FR7}
\end{table}
In general does availability in the context of \ac{IIoT} describe guarantees, that system resources are always available \cite{Tan_2021}. The most common form of attacks targeting the availability security objective are flooding attacks, including \ac{DoS}, which cause service unavailability by generating traffic volume overloading the network capabilities \cite{Pal_2021}. The first \ac{SR} 7.1 requires systems to provide counter measures against such attacks in general. In addition there is the need for communication load management and for the possibility to limit the effect of the \ac{DoS} on other systems with higher \ac{SL}s. The protection against \ac{DoS} in \ac{IIoT} is mainly achieved due to securing the system borders with intrusion prevention systems (IPS), different kinds of firewalls or \ac{DoS} defence systems.\\ As described is the whole compliance monitoring of \ac{DoS} protection not possible due to the hardware-based measures implementing the \ac{RE}s in higher SL-Ts. In general is availability in \ac{IIoT} systems mostly achieved by physical redundancy, which can lead to the assumption that compliance monitoring is much more difficult and system specific \cite{Iwanicki_2018}. This fact realizes within the monitorable attributes for the SRs 7.2 to SR 7.6, as they form the physical representation of the availability requirements. Because of that have manual attributes to be set before the process is executed and then checked by the \ac{CMS}, noted that there could be logical or traffic attributes in specific scenarios. An example for such a hardware-based \ac{SR} is provided by SR 7.5 describing the necessity for an emergency power setup.\\ The SR 7.6 requires the control system to provide and adapt a collection of network and security configuration settings in general and on SL three and higher a machine-readable version. This requires a logical or manual attribute stating if such a configuration is existent and can be pulled. If the configuration is machine readable, the attribute becomes a logical attribute, as the compliance system can actively pull the security configurations or has access to these configurations from the beginning. Otherwise has the attribute to be set manually. SR 7.7 requires the \ac{IACS} to fulfill the common security principle of least functionality. In the context of the IEC 62433 requirement does this mean, that the system must restrict the usage of unnecessary functions, ports, protocols, and services \cite{Kobes2021}. As this consists of multiple indicators being combined is a logical binary attribute necessary to check the overall compliance status to SR 7.7. To set this attribute can multiple lists of expected behavior be used to match with monitored traffic. For example can a list of expected protocols be used to check if a monitored protocol fulfills the intended \ac{SL} of the system. If a communication with an unexpected protocol is monitored, it can be assumed that compliance to the least privilege principle is not given. The last \ac{SR} in this \ac{FR} requires the \ac{IACS} to provide a list of all the installed components connected to the control system. Even though such lists can be used to measure other attributes, can the existence of the functionality to provide such a list only be represented by a logical or manual binary attribute.

\section{Conclusion}
The market study provided an extensive insight on the structure of monitorable attributes in IIoT systems. It defined different types of attributes, conducted an investigation of monitorable properties and summarized them. Bases on these attributes it is possible to develop an approach to security standard compliance monitoring. A system in that scope would assign these attributes to a collection of \acp{SR} and regularly check them to generate a compliance report.

\phantomsection
\addcontentsline{toc}{chapter}{Bibliography}
\bibliographystyle{alpha}
\bibliography{References}

\end{document}